\def\PRL{{ Phys. Rev. Lett.\ }\/}
\def\PRB{{ Phys. Rev. B\ }\/}

\def\be{\begin {equation}}
\def\ee{\end {equation}}
\def\ber{\begin {eqnarray}}
\def\eer{\end {eqnarray}}
\def\bers{\begin {eqnarray*}}
\def\eers{\end {eqnarray*}}

\makeatletter

\newcommand{\Rmnum}[1]{\expandafter\@slowromancap\romannumeral #1@}
\makeatother

\makeatletter
\newcommand*\env@matrix[1][*\c@MaxMatrixCols c]{%
  \hskip -\arraycolsep
  \let\@ifnextchar\new@ifnextchar
  \array{#1}}
\makeatother

\documentclass[aps,prb,showpacs,superscriptaddress,a4paper,twocolumn]{revtex4-1}

\usepackage{amsmath}
\usepackage{float}
\usepackage{color}

\usepackage[normalem]{ulem}
\usepackage{soul}
\usepackage{amssymb}

\usepackage{amsfonts}
\usepackage{amssymb}
\usepackage{graphicx}
\begin {document}

\title{Coexistence of Multifold  and Multidimensional Topological  Phonons in KMgBO$_{3}$ } 

\author{P. C. Sreeparvathy}
\affiliation{Department of Physics, Indian Institute of Technology Bombay, Powai, Mumbai 400076, India}

\author{Chiranjit Mondal}
\affiliation{Department of Physics, Indian Institute of Technology Bombay, Powai, Mumbai 400076, India}
\affiliation{Center for Correlated Electron Systems, Institute for Basic Science (IBS), Seoul 08826, Korea}

\author{Chanchal K. Barman}
\affiliation{Department of Physics, Indian Institute of Technology Bombay, Powai, Mumbai 400076, India}
\affiliation{Department of Physics, Sungkyunkwan University, Suwon 16419, Republic of Korea}

\author{Aftab Alam}
\email{aftab@iitb.ac.in}
\affiliation{Department of Physics, Indian Institute of Technology Bombay, Powai, Mumbai 400076, India}

\date{\today}

\begin{abstract}
Topological interpretations of phonons facilitate a new platform for novel concepts in phonon physics. Though there are ubiquitous set of reports on topological electronic excitations, the same for phonons are extremely limited. Here, we propose a new candidate material, KMgBO$_{3}$, which showcase the co-existence of several multifold and multidimensional topological phonon excitations, which are protected by spatial and non-spatial symmetries. This includes zero dimensional double, triple and quadratic Weyl phonon nodes, one dimensional nodal line/loop and two dimensional doubly degenerate nodal surface states. Nodal line/loop emerges from the spin-$\frac{1}{2}$ phonon nodes, while the two dimensional doubly degenerate nodal surface arises from a combination of two fold screw rotational and time reversal symmetries. Application of strain breaks the C$_3$ rotational symmetry, which annihilates the  spin-1 double Weyl nodes, but preserves other topological features. Interestingly, strain helps to create two extra single Weyl nodes, which in turn preserve the total chirality. Alloying also breaks certain symmetries, destroying most of the topological phonon features in the present case. Thus, KMgBO$_{3}$ is a promising candidate which hosts various Weyl points, large Fermi arcs with a very clean phonon spectra and tunable topological phonon excitations, and hence certainly worth for future theoretical/experimental investigation of topological phononics.
 \end{abstract}

\maketitle

{\par}{\it Introduction:} Topological properties of fermionic states enforced by spatial and non-spatial symmetries have been ubiquitously explored in the literature.\cite{Kaustuv,YXia,ShuoWang,HaoZheng,MasatoshiSato} The same for bosons (photons, phonons), however, has started only recently\cite{Sebastian,Roman,Emil} and there are only limited studies along these lines. The concept of chiral phonons is further extended to materials with non-symmorphic crystalline symmetry, which unlocks the applications in THz frequency range.\cite{Tiantian,Jiangxu,JiangxuLi} In addition, topological phonons also find applications in unconventional heat transfer, electron phonon coupling, and  phonon diode.\cite{Jiangxu, Emil,Yizhou} The advantage of topological bosons over fermions is the possibility of probing topological states in any frequency range using experimental techniques unlike fermionic states, where this is mostly possible only near the Fermi level (E$_F$). In line with electronic states, topological phonons also show anomalous transport behaviour, such as anomalous phonon hall effect.\cite{Lifa,KangtaiSun,TaoQin} Over the years, diverse quasiparticle excitations have emerged as a consequence of different band crossings which can host non-zero topological Chern number. For instance, conventional spin-$\frac{1}{2}$ Weyl fermions possess Chern number $\pm$1. Moreover, unconventional chiral particles with Chern number greater than unity have also been realised in crystalline systems. Threefold spin-1 and fourfold charge-2 Dirac fermions are examples of such higher Chern number quasiparticles.\cite{Zhicheng,ShiXinZhang,chanchal} These multifold band crossings yield a zero dimensional (0D) point degeneracy in momentum space.\cite{MZahidHasan} There exists other types of topological band crossings such as nodal line/loop and nodal surface which induce one dimensional(1D) and two dimensional (2D) degeneracy respectively.\cite{XiaotianWang}

A proper understanding and hunt for real materials which host multiple topological phonons have emerged as a frontline area of research due to its demand from both fundamental and application point of view. Several 3-dimensional and 2-dimensional materials are investigated for topological phonons.\cite{RYWang,ChengwuXie,YuanjunJin} Coexistence of a few multidimensional topological phonons have been identified in some materials.\cite{Jianhua} However, to the best of our knowledge, coexistence of multifold and multidimensional (such as nodal points(NP), nodal line(NL) and nodal surfaces(NS)) phononic quasiparticles is never reported in a real material.

\par
Tuning the topological properties of a material via strain/pressure/doping is a well accepted method which helps to modify the symmetry and hence the electronic structure, facilitating the onset of trivial/nontrivial topological excitations.\cite{CJEklund,LeiJin} There has been several theoretical/experimental studies which has showcased such symmetry mediated transition from  topological insulator to semimetal and vice versa, creation and annihilation of Weyl/Dirac and other kinds of topological nodes.\cite{ChunLin,SteveM,YanSun} A similar band engineering and the associated transition can be expected for phonon quasiparticles as well.
\par
In this letter, we present the co-existence of several topological phonon excitations in a single compound, KMgBO$_{3}$, which crystallizes in a cubic phase.  KMgBO$_{3}$  is an experimentally synthesised  chiral compound, which has been explored for nonlinear optical properties.\cite{exp_paper,JianghuiZheng} Recently, this compound is also found to show interesting topological electronic properties by us.\cite{chanchal}  One of the main focus of the present study is to investigate the topological phonon excitations and the effect of symmetry breaking on topological properties of this material. Pristine KMgBO$_{3}$ has a rich phonon band dispersion, showing several topologically non-trivial multifold/multidimensional band crossings in different frequency range. This includes spin-1 Weyl, charge 2 Dirac, single Weyl, nodal line, nodal loop and nodal surface states. These states are topologically protected by crystalline and time reversal symmetries. Application of strain breaks the C$_3$ rotational symmetry, which annihilates the spin-1 double Weyl nodes but preserve the charge 2 Dirac node and nodal surface states.
We have further disturbed the symmetry by doping with Rb, transforming it to a triclinic structure. Though doping preserves the total chirality, it destroys all the topological phonon features. Additionally, we have explored two other prototype compounds RbMgBO$_{3}$\cite{RVKurbatov} and CsCdBO$_{3}$,\cite{HongweiYu} which show similar topological features.
\begin{figure}[t]
\centering
         \includegraphics[width=1.02\linewidth]{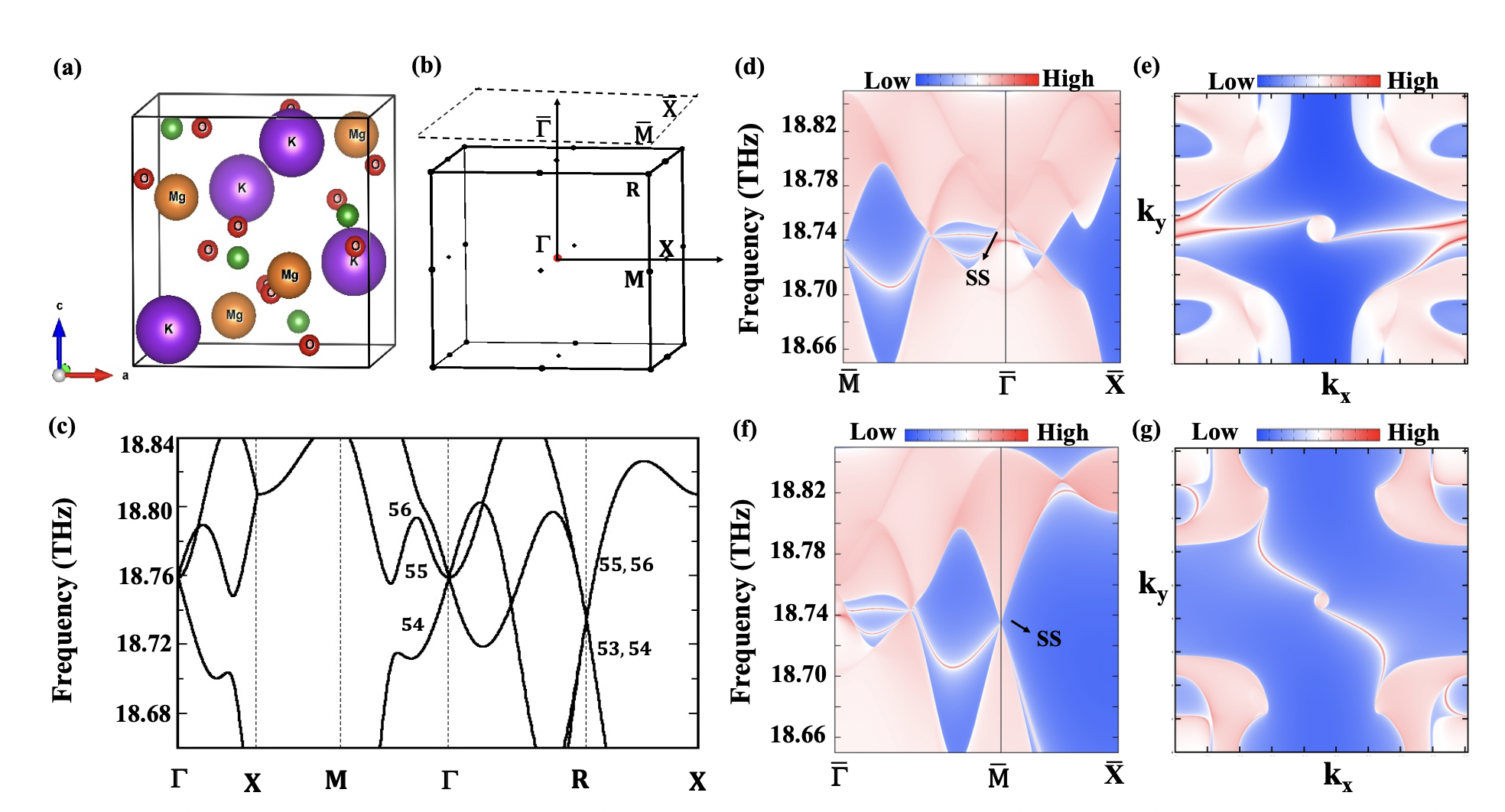}
\caption{(Color online) For KMgBO$_3$, (a) Crystal structure (space group P2$_{1}$3(\# 198)) (b)  Brillouin zone(BZ) for bulk and (001) surface (c) Phonon dispersion in a small frequency interval where a 3-fold spin -1 Weyl and 4-fold charge 2 Dirac point is observed. (d,f) Surface state around $\bar{\Gamma}$ and $\bar{M}$ point on (001) surface, corresponding to the Spin-1 Weyl node and charge 2 Dirac node respectively. (e,g) Surface arc at 18.76 THz and 18.73 THz between $\bar{\Gamma}$ and $\bar{M}$ points.}
\label{fig1}
\end{figure}

{\par}{\it Computational Details:} First-principles calculations are done using the
Vienna Ab initio Simulation Package (VASP)\cite{Kresse1,Kresse2} and Phonopy.\cite{Togo} Other computational details are provided in the supplementary material(SM).\cite{suppl}
\par 
{\it  Crystal Structure :} KMgBO$_{3}$ crystallizes in cubic structure with space group P2$_{1}$3 (\#198).\cite{exp_paper} The theoretically optimised lattice parameter (6.09 $\AA$) matches fairly well with the experiment (6.8345 $\AA$).\cite{exp_paper}  Figure \ref{fig1}(a,b) show the real space crystal structure and bulk and (001) surface Brillouin zone (BZ). The space group 198 holds the tetrahedron (T4) point group symmetry, which provides two twofold screw rotations, S$_{2z}$ = \{C$_{2z}|\frac{1}{2},0,\frac{1}{2}|$\}, S$_{2y}$ =\{C$_{2z}|0,\frac{1}{2},\frac{1}{2}|$\} and one threefold rotation, S$_{3}$= \{C$^{+}$$_{3,111}$$|$0,0,0\} as generators at $\Gamma$ point. At R point, the generators are S$_{2x}$ = \{C$_{2x}|\frac{1}{2},\frac{3}{2},0|$\}, S$_{2y}$ = \{C$_{2y}|0,\frac{3}{2},\frac{1}{2}|$\} and S$_{3}$= \{C$^{-}$$_{3,111}$$|$0,0,0\}, while at X point, they are  S$_{2y}$ = \{C$_{2y}|0,\frac{1}{2},\frac{1}{2}|$\} and S$_{2z}$ = \{C$_{2z}|\frac{1}{2},0,\frac{1}{2}|$\}.\cite{book} In addition, the space group preserves time-reversal symmetry($T$).
\par
{\it  Phonon spectra:} KMgBO$_{3}$ structure involves four formula units per unit cell leading to 24 atoms in the cell, which generate 3 acoustic and 69 optical phonon modes. The absence of imaginary modes confirms the dynamical stability of the compound. Phonon dispersion over a wider frequency range (0-40 THz) is presented in Fig. S1(a).\cite{suppl} A close inspection of bulk phonon dispersion reveals the prosperity of different nature of band crossings in different frequency range, yielding topological Weyl, nodal line/loop and nodal surface states, as detailed below. We have also studied the surface phonon spectra.
\begin{figure*}[t]
\centering
\includegraphics[width=0.9\linewidth]{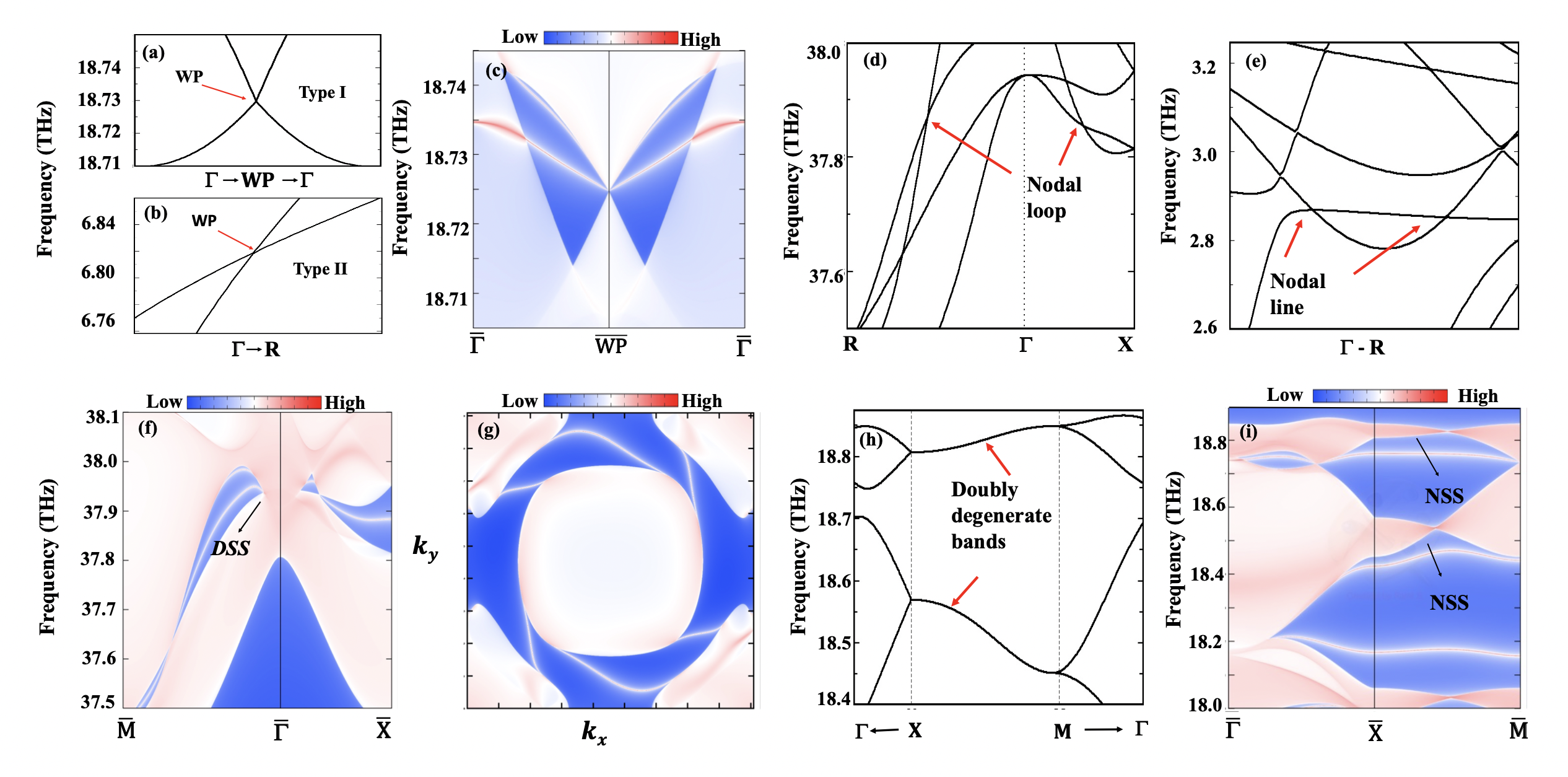} \\
\caption{(Color online) For KMgBO$_3$, the phonon bands along (a) $\Gamma$-X and (b) $\Gamma$-R, showing type-I and type II Weyl crossings respectively (c) phonon surface spectra on (001) surface corresponding to type-I Weyl node (d) phonon dispersion along R-$\Gamma$-X confirming a nodal loop behavior (e) phonon dispersion along $\Gamma$-R confirming a nodal line behavior (f) drumhead like surface state on (001) surface corresponding to the nodal loop shown above (g) surface arc due to the nodal loop on k$_{x}$-k$_{y}$ plane (h) phonon dispersion along X-M, showing the doubly degenerate bands (i) surface spectra due to the doubly degenerate bands.}
\label{fig2}
\end{figure*}
\par
{\it  Double Weyl Phonons :} A symmetry enforced 3- or 4-fold band crossings can induce unconventional topological excitations.\cite{BBradlyn} Let us first consider the case at $\Gamma$ point. For phonons, the square of the time-reversal symmetry operator($T$) is identity($I$). The generators S$_{2z}$ and S$_{2y}$ commute each other at $\Gamma$ point, which can be understood by considering the transformation of lattice coordinates as a function of the twofold screws C$_{2z}$ and C$_{2y}$.  These transformations are shown in Ref. \onlinecite{footnote0}.

To understand the degeneracy of bands at $\Gamma$,  one need to consider the eigenstates generated by these symmetry operators which commute with the Hamiltonian of the system,
S$_{2z}$$|\psi>$ = $\lambda$$_{1}$$|\psi>$, S$_{2y}$$|\psi>$ = $\lambda$$_{2}$$|\psi>$, where
 $\lambda$$_{1}$ =$\pm$1, $\lambda$$_{2}$ =$\pm$1 according to S$^2$$_{2z}$ = 1 and S$^2$$_{2y}$ = 1.
S$_{3}$ obeys S$_{2z}$S$_{3}$=S$_{3}$S$_{2y}$ and S$_{3}$S$_{2z}$S$_{2y}$ = S$_{2y}$S$_{3}$,  which impose a threefold degeneracy at $\Gamma$ with eigenstates $|\psi>$, $|$S$_{3}$$\psi$$>$ and $|$S$^{2}$$_{3}$$\psi$$>$, when screws are nontrivial.\cite{chanchal} When screws are trivial, the $\Gamma$ point can hold either nondgenerate or twofold degeneracy.\cite{chanchal} The threefold degenerate bands at $\Gamma$ arises from a combination of two highly linearised bands and one flat band, which can be defined by spin-1 states with chirality +2,0 and -2.\cite{BBradlyn,chanchal}
Two screws S$_{2x}$ and S$_{2y}$ anticommute each other at R point, and  S$^2$$_{2x}$ = -1, S$^2$$_{2y}$ = -1. This induces a twofold degenerate state (with eigenvalue $\pm$i) at R point. Further, the combination of threefold rotation S$_{3}$ and twofold screws create two more distinct states, enabling a fourfold charge 2 Dirac point with chirality $-2$,$-2$, $+2$ and $+2$.  Further symmetry related details can be found in our previous study  performed for electronic excitation.\cite{chanchal}

Figure \ref{fig1}(c) displays the phonon dispersion in a smaller frequency range where a threefold spin-1 Weyl point (at $\Gamma$) and a fourfold charge 2 Dirac point (at R) are observed. At $\Gamma$ point, near 18.76 THz, three bands (band number 54, 55 and 56) cross degenerately, with Chern number -2 (for band 54), 0 (for 55) and +2 (for 56).  At R point, near 18.73 THz, a fourfold degenerate band crossing arises which involve 53, 54, 55 and 56$^{th}$ bands (Fig. \ref{fig1}(c)). This fourfold unconventional Dirac like node can be seen as a combination of two identical spin- $\frac{1}{2}$ Weyl phonons, with a net topological charge 2.\cite{BBradlyn} A similar spin-1 Weyl and charge $2$ Dirac node also occur at other frequency range, one of which is shown in Fig. S1(b) of SM.\cite{suppl} Simulated Berry curvature  at $\Gamma$ point (for -2 chirality) in the k$_{x}$-k$_{y}$ plane is shown in Fig. S1(c),\cite{suppl} clearly confirming the flow of Berry flux from $\Gamma$ point. Due to the large separation between bulk double Weyl phonons, a clean surface spectra is expected as analysed below.

 \par 
Surface states of the double Weyl phonons are examined on (001) and (111) surfaces. Figure \ref{fig1}(d) and \ref{fig1}(f) show the (001) surface state corresponding to spin-1 Weyl point and charge $2$ Dirac point projected on $\bar{\Gamma}$  and $\bar{M}$ points respectively. Two linearly dispersed and one parabolic surface state can be seen at  $\bar{\Gamma}$ point, which are originated from three bulk bands with chirality $-2$ ,$0$,$2$, while a highly linearised surface state is observed at $\bar{M}$ point, which represent the charge 2 Dirac point. Since these two nodes at $\bar{\Gamma}$ and $\bar{M}$ possess opposite chirality, one can expect a surface arc connecting these two points. A surface arc simulated at 18.76 THz is shown in Fig. \ref{fig1}(e), which indirectly connects $\bar{\Gamma}$ and $\bar{M}$ points. Figure \ref{fig1}(g) shows the surface arc projected at a slightly lower frequency (18.73 THz) which directly connects the two points, with a long diagonal. Such large surface arcs are much easier to probe experimentally, as reported earlier for other  topological phonon materials.\cite{Tiantian} Further, the analysis of (111) surface shows the presence of charge 2 Dirac node, see Fig. S1(d) of SM.\cite{suppl} The surface state of spin-1 Weyl point located at higher phonon frequency is also presented in SM\cite{suppl} (see Fig. S1(e)).


{\par}{\it Single Weyl Phonons :} Pristine KMgBO$_{3}$ also hosts two types of spin $\frac{1}{2}$ topological states, (1) zero dimensional Weyl points and (2) one dimensional Nodal line/Nodal ring. First let us discuss the single Weyl point. Weyl nodes can be formed by breaking either the time-reversal symmetry or the inversion symmetry. In the present case, it arises from the noncentrosymmetric nature of KMgBO$_{3}$.  Weyl nodes can be further classified into type I and II, according to the nature of band crossings.\cite{footnote1,Alexey,BWXia} In KMgBO$_{3}$ phonon spectra, both types of Weyl nodes are observed, as shown in Fig. \ref{fig2}(a,b). Type I is observed on  '$xy$' plane in the frequency range 18.71-18.74 THz (formed by the crossing of 54$^{th}$ and 55$^{th}$ phonon bands). While, type II is observed along $\Gamma$-R line in the frequency range 6.76-6.84 THz (formed by the crossing of 28$^{th}$ and 29$^{th}$ bands). We have chosen one of the Weyl points (type-I) to present its surface state on (001) surface, as shown in Fig. \ref{fig2}(c). Clearly, this is a highly linearised surface state with a pristine surface arc.

Next, we will discuss two other types of spin-$\frac{1}{2}$ topological nodal points.
Presence of more than one nodal points at the same frequency but different momentum forms one dimensional nodal line/loop like band crossing. Figure \ref{fig2}(d) shows one such  example where two tilted nodal points are observed along  two different high symmetry lines ($\Gamma$-X and $\Gamma$-R) at the same frequency. This causes the formation of a nodal loop. Similarly, there exists nodal loops along other directions such as $\Gamma$-Y and $\Gamma$-R, $\Gamma$-Z and $\Gamma$-R. Additionally, the phonon dispersion along $\Gamma$-R show two nodal points at same frequency which provides a nodal line behaviour (see Fig. \ref{fig2}(e)). Since the bulk R point projects as $\bar{M}$ on (001) surface, it gives a drumhead-like surface due to the nodal loop around $\bar{\Gamma}$ point. This is shown in Fig. \ref{fig2}(f) and the corresponding iso-surface arc in Fig. \ref{fig2}(g). A similar nature of iso-surface has been observed in a carbon allotrope.\cite{JingYangYou}

{\par}{\it  {Nodal Surface} :} Higher dimensional band degeneracy has been theoretically predicted\cite{QingBo} and experimentally realised in recent past.\cite{Yihao} There are few symmetry mediated conditions which impose such high dimensional band degeneracy.\cite{footnote2} The bands along X-M and X-R aligned in k$_{x}$=$\pi$ plane provide a minimum of double degeneracy due to the above condition.\cite{ChengwuXie,QingBo} Figure \ref{fig2}(h) shows the double degenerate band along X-M. Similarly, a doubly degenerate phonon band can be seen in k$_{y,z}$=$\pi$ planes. It is evident from the figure that, these doubly degenerate bands are well separated and hence can be easily realised experimentally.
 Figure \ref{fig2}(i) shows the corresponding surface states, which is mostly flat with a low spread in frequency range enabling KMbBO$_{3}$ to be a promising candidate for ideal nodal surface.  In a similar way, one can observe a nodal surface on other two surfaces.  A combined visual of  these nodal surface create and nodal cube around the BZ centre. This is a unique surface feature of KMgBO$_{3}$ which stands out in comparison to other reported topological nodal surface materials.\cite{Xiaotian}

\begin{figure}[t]
\centering
\includegraphics[width=\linewidth]{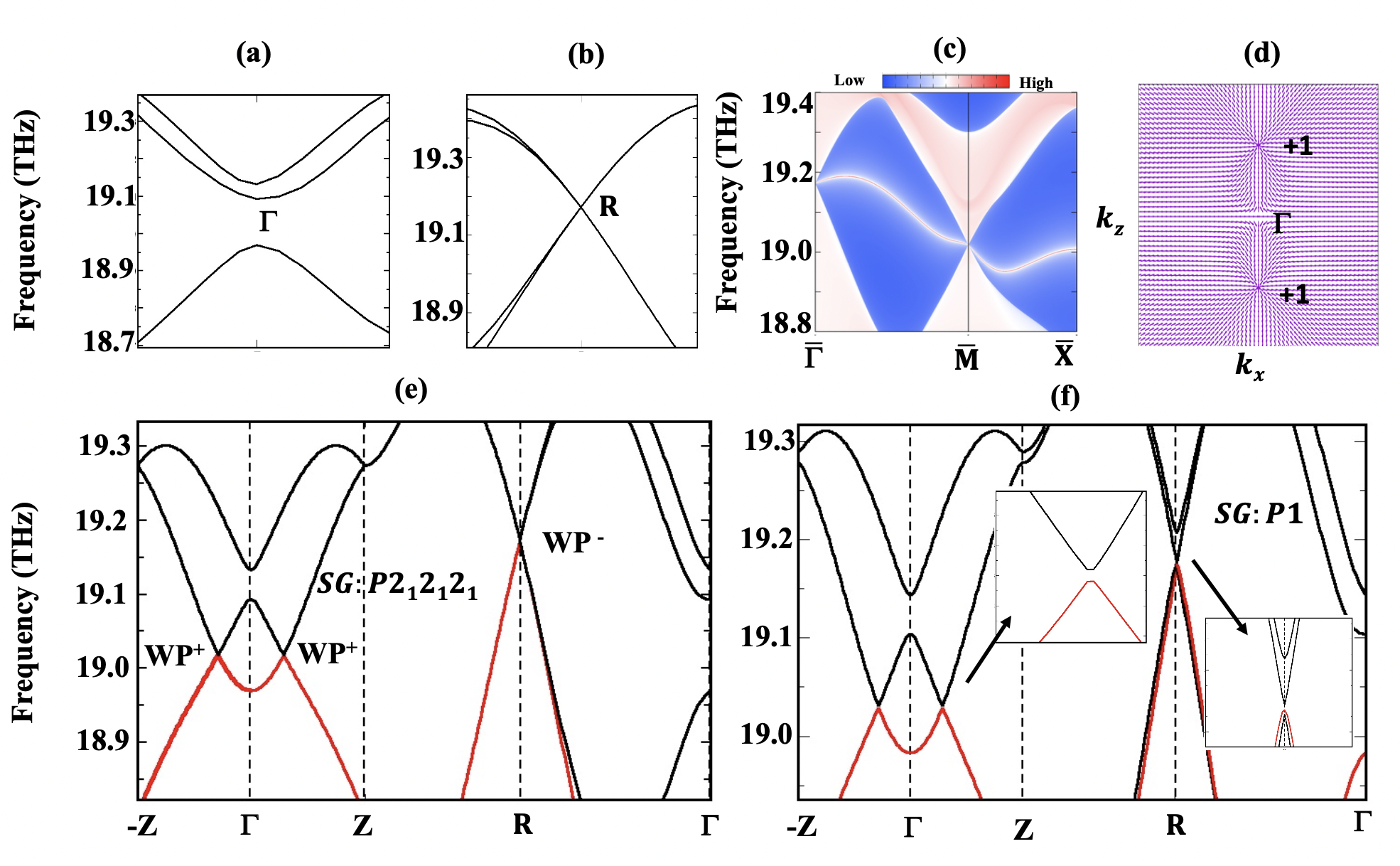}

\caption{(Color online) For KMgBO$_{3}$ under 3$\%$ compressive strain: phonon bands (a) around $\Gamma$ point and (b) R point. (c) Surface state corresponding to charge 2 Dirac node at R point. (d) Berry curvature at two Weyl points on k$_{x}$-k$_{z}$ plane. (e) Phonon dispersion in strained state (space group P2$_{1}$2$_{1}$2$_{1}$), representing the charge 2 Dirac point at R (with chirality -2) and two Weyl nodes (with chirality +1) along -Z - $\Gamma$-Z. (f) Phonon dispersion for Rb-substituted KMgBO$_3$ (space group P1), representing the annihilation of charge 2 Dirac point at R and the Weyl nodes along -Z - $\Gamma$-Z  direction.}
\label{fig3}
\end{figure}

{\par}{\it {Strain effects} :}  Both uni-axial (along $c$ axis) and bi-axial  (along $a$ and $b$ axes) compressive and tensile strains are applied on KMgBO$_{3}$ to illustrate the effect of symmetry breaking on topological phonon excitations. Such strain breaks the cubic symmetry and transforms the structure to an orthorhombic space group P2$_{1}$2$_{1}$2$_{1}$ (\# 19), which lacks the threefold C$_{3}$ rotational symmetry\cite{book} along with deltahedron type (D$^{4}$$_{2}$) point group (see footnote\cite{footnote3} for more symmetry related details).
\begin{figure}[t]
\centering
\includegraphics[width=\linewidth]{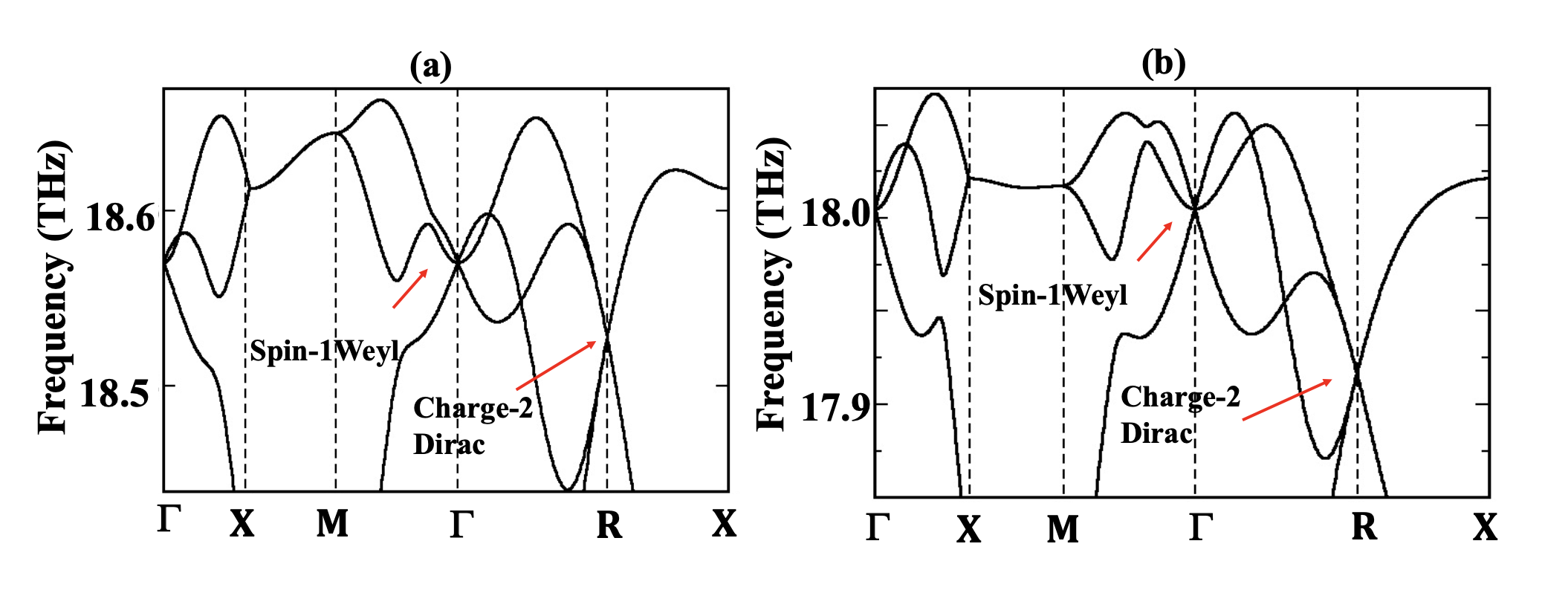}
\caption{(Color online) Phonon dispersion for (a) RbMgBO$_{3}$ and (b) CsCdBO$_{3}$.}
\label{fig4}
\end{figure}

\par
Let us now understand the emergence of different topological phonon excitations as a function of uni-axial strain. Here we applied a maximum of 3$\%$ strain (both in compressive and tensile forms)  in an interval of 1$\%$  along $c$ axis. At ambient state (cubic phase), KMgBO$_{3}$ hosts two kinds of double Weyl phonons i.e. spin-1 type at $\Gamma$ and charge 2 Dirac type at R points. It is clear that, in strained system, the generators at $\Gamma$ will enforce the bands to become non-degenerate (since the two screw operators commute), which should destroy the spin-1 type double Weyl phonon nature. The eigenvalues can be $\pm$1 at $\Gamma$ point.\cite{RMatthias} This is precisely shown in Fig. \ref{fig3}(a) (under 3$\%$ compressive strain). Coming to R point, the possible combinations of eigenvalues are ($1$,-$1$,-$1$), (-$1$,$1$,-$1$), (-$1$,-$1$,$1$) and ($1$,$1$,$1$)\cite{RMatthias} which provide a fourfold degeneracy and assure the possibility of preserving the charge 2 Dirac point.  The bulk phonon dispersion at R point is shown in Fig. \ref{fig3}(b)(under 3$\%$ compressive strain), confirming the persistence of charge 2 Dirac state at R. To confirm the double Weyl phonon nature, the surface state of charge 2 Dirac point is simulated, which shown in Fig. \ref{fig3}(c). A highly linearised surface state along with a clear surface arc is clearly visible from this figure.
\par
Further, let us analyse the effect of strain on spin-$\frac{1}{2}$  Weyl phonons. The analysis of band topology between 54$^{th}$ and 55$^{th}$ bands shows the presence of two Weyl points  with +1 chirality around $\Gamma$-point (with locations (0,0,+k$_z$) and (0,0,-k$_z$)). This balances the chirality generated due the double Weyl phonons at R point. Figure \ref{fig3}(d) displays the computed Berry curvature on k$_{x}$-k$_{z}$ plane, which shows the outcome as expected. Importantly, the combination of double and single Weyl phonons conserves the Ninomiya theorem\cite{HBNielsen} in strained state. To further understand this point, we have disturbed the symmetry by substituting one of the K element by Rb. This transforms the system to a low symmetry triclinic space group (P1), which lacks all the screw rotation symmetries. The phonon dispersion of strained (P2$_{1}$2$_{1}$2$_{1}$) and the doped phase (P1) are presented in Fig. \ref{fig3}(e) and \ref{fig3}(f) respectively.  The Dirac nature at R point is defaced in the P1 phase, which arises purely due to the breaking of screw symmetry, with the simultaneous defacing of +1 chiral Weyl points along  -Z- $\Gamma$-Z direction.
 \par
Further, the `1D' nodal ring/loop is destroyed under uni-axial strain, with the opening of a gap along $\Gamma$-R. Regarding the `2D' nodal surface, the presence of screws and time-reversal symmetry induce a Kramers like doubly degeneracy at k$_{i}$=$\pi$ which cause the formation of nodal surface. The application of biaxial strain alters the topological phonon excitations in a similar way. The phonon dispersion around the $\Gamma$ and R point under 3$\%$ bi-axial strain are shown in SM.\cite{suppl} Changes in other topological features remain in line with those of uni-axial strain.

{\par}{\it {Other materials} :} We have also analysed the topological phonon excitations of two other experimentally synthesized prototype compounds RbMgBO$_{3}$ and CsCdBO$_{3}$.  Optimised lattice parameter for RbMgBO$_{3}$ and CsCdBO$_{3}$ are 7.02 $\AA$ and  7.629 $\AA$ respectively, which are in good agreement with experiment.\cite{RVKurbatov,HongweiYu} A similar nature of phonon dispersion is observed for these two compounds, see Fig. \ref{fig4}(a,b). They show various topological phonons, including double Weyl, single Weyl, nodal line/loop and nodal surface. Full phonon dispersion in larger frequency interval and bulk/surface states of spin-1 type double Weyl nodes are shown in SM.\cite{suppl}

{\par}{\it {Summary} :}  We propose three promising candidate materials (KMgBO$_3$, RbMgBO$_3$, and CsCdBO$_3$) for topological phononic applications in different THz frequency range. At ambient condition, these materials provide a rich platform to simultaneously host several multifold and multidimensional topological Weyl phonons, such as spin-1 Weyl, charge 2 Dirac, single Weyl, nodal line/loop and nodal surface. These topological states are protected by crystalline and time reversal symmetries. Application of strain breaks the C$_{3}$ rotational symmetry which mediates the annihilation of spin-1 double Weyl nodes while preserves the other topological features. Substitution at K/Rb/Cs-site also lowers the symmetry to triclinic structure, and destroys most of the topological features. The present work, for the first time, showcase a novel candidate material where multiple topological phonon excitations coexist and hence can be an interesting platform for future theoretical/experimental investigations.

{\par}{\it  Acknowledgements:}
AA acknowledges DST SERB, India (Grant No. CRG/2019/002050) for funding to support this research.  SPC thank IIT Bombay for institute postdoctoral fellowship and computing facility.C.M. would like to thank Institute for Basic Science in Korea (Grant No. IBS-R009-D1) for funding.


\begin{thebibliography}{113}
\bibitem{footnote0} The transformations are C$_{2z}$($x$,$y$,$z$) $\rightarrow$ ($-x$,$-y$,$z$) and C$_{2y}$($x$,$y$,$z$) $\rightarrow$ ($-x$,$y$,$-z$), which define S$^2$$_{2z}$ = 1, S$^2$$_{2y}$ = 1 and S$_{2z}$S$_{2y}$ = S$_{2y}$S$_{2z}$.\cite{book} The lattice coordinate transformation due to C$_{3}$ is C$_{3,111}$($x$,$y$,$z$) = ($z$,$x$,$y$), which shows S$_{3}$$^{2}$ = 1.
\bibitem{footnote1}The general definition of the type I and type II Weyl nodes in a fermionic systems can be understand the following way: The Weyl point, which possess a zero like Fermi surface is defined as type I and the type II shows a tilted dispersion along the momentum space and make an overlap between the electron and hole pockets.\cite{Alexey} A phonon counterpart for both type I and type II can be understand in the similar way.\cite{BWXia}
\bibitem{footnote2}The combination of twofold screw rotational symmetry and the  time-reversal symmetry operators leads to the formation of an anti-unitary operator ($T$$C$$_{2i}$)$^2$ = $-1$, which opens up the possibility of a higher dimensional (2D) band degeneracy at k$_{i}$=$\pi$ plane. As a result of ($T$$C$$_{2i}$)$^2$ = $-1$, at $k$$_{i}$=$\pi$  high symmetry plane the bands act like Kramers degenerate with eigenvalue $\pm$i.
\bibitem{footnote3} Space group P2$_12_12_1$ has Deltahedron type (D$^{4}_2$) point group involving three twofold screw rotation symmetries. At $\Gamma$ point, the generators are S$_{2z}$ = \{C$_{2z}|\frac{1}{2}|,0,\frac{1}{2}|$\} and S$_{2y}$ = \{C$_{2z}|\frac{1}{2}|,\frac{1}{2},0|$\}.  These two satisfy the commutation relation at this point  [S$_{2y}$ ,S$_{2z}$ ] =0.  At R point, the symmetries are S$_{2x}$ = \{C$_{2z}|0,\frac{1}{2},\frac{1}{2}|$\}, S$_{2x}$ = \{C$_{2z}|\frac{1}{2},\frac{1}{2},0|$\}, and identity operator. The two screws are anticommute at R point.  At X point the symmetries are S$_{2x}$ = \{C$_{2x}|0,\frac{1}{2},\frac{1}{2}|$\} and S$_{2z}$ = \{C$_{2x}|\frac{1}{2},0,\frac{1}{2}|$\}
\end{thebibliography}
\end{document}